\documentclass[10pt,journal,compsoc]{IEEEtran}
\usepackage{kotex}
\usepackage{tikz}
\usetikzlibrary{shapes.geometric, arrows.meta, positioning}
\ifCLASSOPTIONcompsoc
  \usepackage[nocompress]{cite}
\else
  \usepackage{cite}
\fi

\ifCLASSINFOpdf
\else
\fi

\usepackage{tabularx}
\usepackage{booktabs}
\usepackage{amsmath}
\usepackage{amssymb}
\DeclareMathOperator*{\argmin}{arg\,min}

\usepackage{enumitem}
\usepackage{makecell}
\usepackage{array}
\usepackage{caption}
\usepackage{booktabs}

\usepackage{graphicx}
\graphicspath{ {./images/} }
\usepackage{subcaption}
\usepackage[export]{adjustbox}
\usepackage{wrapfig}

\usepackage[hidelinks]{hyperref}
\usepackage{authblk}

\begin{document}

\captionsetup[table]{labelformat={default},labelsep=period, name={Table}}

\captionsetup[figure]{labelformat={default},labelsep=period, name={Figure}} %labelfont={bf},

\title{Leveraging Surplus Electricity:\\
Profitability of Bitcoin Mining as a National Strategy in South Korea}

% affiliations
\author{
    \IEEEauthorblockN{
        Yoonseul Choi\textsuperscript{1}, 
        Jaehong Jeong\textsuperscript{2, 3}, 
        Jungsoon Choi\textsuperscript{2, 3}
    }\\
    \IEEEauthorblockA{
        \textsuperscript{1}Department of Applied Statistics, Hanyang University, Seoul, South Korea
    }\\
    \IEEEauthorblockA{
        \textsuperscript{2}Department of Mathematics, Hanyang University, Seoul, South Korea
    }\\
    \IEEEauthorblockA{
        \textsuperscript{3}Research Institute for Natural Sciences, Hanyang University, Seoul, South Korea
    }\\    
    \IEEEauthorblockA{
        Emails: \href{mailto:clodagh@hanyang.ac.kr}{clodagh@hanyang.ac.kr}, 
        \href{mailto:jaehongjeong@hanyang.ac.kr}{jaehongjeong@hanyang.ac.kr}, 
        \href{mailto:jungsoonchoi@hanyang.ac.kr}{jungsoonchoi@hanyang.ac.kr}
    }
}

\markboth{Y. Choi et al.: Leveraging Surplus Electricity: Profitability of Bitcoin Mining as a National Strategy in South Korea}%
{Y. Choi \MakeLowercase{\textit{et al.}}: Feasibility of Surplus Electricity for Bitcoin Mining}

\IEEEtitleabstractindextext{%
\begin{abstract}
This study examines the feasibility and profitability of utilizing surplus electricity for Bitcoin mining. Surplus electricity refers to the remaining electricity after net metering, which can be repurposed for Bitcoin mining to improve Korea Electric Power Corporation's (KEPCO) energy resource efficiency and alleviate its debt challenges. Net metering (or net energy metering) is an electricity billing mechanism that allows consumers who generate some or all of their own electricity to use that electricity when they want, rather than when it is produced. Using the latest Bitcoin miner, the Antminer S21 XP Hyd, the study evaluates daily Bitcoin mining when operating at 30,565 and 45,439 units, incorporating Bitcoin network hash rates to assess profitability.

To examine profitability, the Random Forest Regressor and Long Short-Term Memory models were used to predict the Bitcoin price. The analysis shows that the use of excess electricity for Bitcoin mining not only generates economic revenue, but also minimizes energy loss, reduces debt, and resolves unsettled payment issues for KEPCO.

This study empirically investigates and analyzes the integration of electricity surplus in South Korea with bitcoin mining for the first time. The findings highlight the potential to strengthen the financial stability of KEPCO and demonstrate the feasibility of Bitcoin mining. In addition, this research serves as a foundational resource for future advancements in the Bitcoin mining industry and the efficient use of energy resources.
\end{abstract}

}

\maketitle
\IEEEdisplaynontitleabstractindextext
\IEEEpeerreviewmaketitle

\ifCLASSOPTIONcompsoc
\IEEEraisesectionheading{\section{Introduction}\label{sec:introduction}}
\else
\section{Introduction}
\label{sec:introduction}
\fi

\IEEEPARstart{k}{orea} Electric Power Corporation (KEPCO)\footnote{55, Jeollyeok-ro, Naju-si, Jeollanam-do, Republic of Korea, available at \url{https://home.kepco.co.kr/kepco/EN/main.do}} is South Korea’s largest and public state-owned electric utility, playing a pivotal role in the generation, transmission, and distribution of electricity, as well as in the development of various energy projects including nuclear, wind, and coal power. Despite its central position in the nation’s energy infrastructure, KEPCO is currently grappling with severe financial challenges. At the end of 2024, KEPCO's total debt on a consolidated basis was KRW 205.18 trillion, up KRW 273.1 billion from the previous year, the highest ever \cite{Hong2025}. Structural issues, such as policy-driven electricity price freezes amid rising energy costs and consumption, have exacerbated this financial strain, potentially leading to future electricity price increases that would unduly burden households. 

In an effort to address these challenges, KEPCO has initiated a project on Jeju island that leverages electricity generated from household solar power systems \cite{Lee2023}. Under the prevailing net metering scheme, these systems supply energy to households while offsetting consumption; however, a significant portion of the generated power remains unaccounted for—referred to here as surplus power. Despite the efficient integration of household solar energy into the grid, this leftover electricity persists and represents an untapped resource. Harnessing this surplus energy to create an independent revenue stream forms the cornerstone of the present study.

In this context, Bitcoin (BTC) emerges as a promising candidate for such an innovative strategy. Introduced in 2009 by Satoshi Nakamoto, Bitcoin has evolved into a cornerstone of the digital economy, valued for its decentralized structure and transparency \cite{nakamoto2008bitcoin}, \cite{RIEDL2024100132}. Over time, Bitcoin has earned the moniker “digital gold” and is increasingly viewed as a strategic asset by governments worldwide \cite{Myung2024}. For instance, reports indicate that the U.S. government is considering the acquisition of up to one million Bitcoins to help manage national debt, while other nations, including China and Russia, are also exploring Bitcoin reserves \cite{Newsis2024}. Table~\ref{tab:bitcoin-holdings} summarizes the Bitcoin holdings of various countries as of December 2024. Bhutan, El Salvador, and Bulgaria, for example, have taken proactive steps to mine or acquire Bitcoin under government initiatives \cite{Won2024}. Bhutan, which began mining Bitcoin in 2017 when its price was around \$5,000, currently holds 12,568 BTC \cite{AIReporter2024}. El Salvador, which adopted cryptocurrency as legal tender, holds over 5,930 BTC. These international examples reflect a growing international trend towards leveraging Bitcoin as part of national economic strategies.

\begin{table}[ht]
\centering
\caption{Bitcoin Holdings by Country (December 2024)}
\label{tab:bitcoin-holdings}
\renewcommand{\arraystretch}{1.2}
\setlength{\tabcolsep}{9pt}
\begin{tabular}{l r r}
\toprule
\textbf{Country} & \textbf{Holdings (BTC)} & \textbf{Estimated Value (\$)} \\ \midrule
Bhutan           & 12,206                 & 1,177,756,940 \\ 
Bulgaria         & 213,519                & 20,602,448,310 \\ 
El Salvador      & 5,930                  & 572,185,700 \\ 
North Korea      & 200,000 (est.)         & 19,298,000,000 \\ 
Ukraine          & 4,600 (est.)           & 443,854,000 \\ 
United States    & 213,000                & 20,552,370,000 \\ 
\bottomrule
\end{tabular}
\begin{flushleft}
\footnotesize \textit{Note}: Bitcoin prices are based on the exchange rate as of December 24, 2024, using \url{https://www.binance.com/en/price/bitcoin}.
\end{flushleft}
\end{table}

Motivated by these developments, the present study investigates the economic feasibility of utilizing the surplus electricity—originally generated from household solar power systems under net metering and left unused—for Bitcoin mining. We use machine learning models, including Long Short-Term Memory networks and Random Forest Regressors, to compare and evaluate profitability under various operational scenarios. This approach enables us to assess the potential revenue outcomes associated with different market conditions and operational strategies, ultimately providing valuable insights into the viability of repurposing surplus solar electricity as a means to alleviate KEPCO’s substantial debt burden.

\subsection{Related Works}
The prediction of Bitcoin prices has been a subject of extensive study in recent years, with a range of methodologies yielding diverse outcomes. Machine learning algorithms, such as Random Forest (RF) and Long Short-Term Memory (LSTM) networks, have been particularly prominent in this field due to their ability to handle nonlinear relationships and time series data.

An and Oh (2021) demonstrated the potential of LSTM models to provide actionable insights in volatile cryptocurrency markets. Their study employed a rigorous methodology, splitting the data into training (444 days), validation (144 days), and testing sets (144 days) to evaluate model performance \cite{an2021}.

Hamayel and Owda (2021) explored the use of advanced machine learning techniques, including LSTM, Bidirectional LSTM (bi-LSTM), and Gated Recurrent Unit (GRU), for predicting the prices of cryptocurrencies such as Bitcoin, Ethereum, and Litecoin \cite{hamayel2021novel}. Their findings revealed that the GRU model outperformed the other algorithms, achieving the highest prediction accuracy with minimal error margins.

Pabuccu et al. (2023) applied various machine learning and deep learning models, such as Support Vector Machines (SVM), Random Forest (RF), and Naïve Bayes (NB), to analyze Bitcoin price fluctuations. The study found that RF excelled in predicting continuous datasets, effectively capturing interactions among variables like Simple Moving Average (SMA) and Momentum. Conversely, NB performed poorly, and artificial neural networks showed superior performance on discrete datasets. Autoregressive integrated moving average (ARIMA) was also highlighted as a computationally efficient option for time series forecasting \cite{pabucccu2023forecasting}.

Azari (2019) compared the performance of ARIMA and Recurrent neural network (RNN)-based models in Bitcoin price prediction. ARIMA was found to be effective for short-term predictions with low volatility but struggled under highly volatile conditions, where LSTM models were better suited for long-term forecasting. The study used evaluation metrics like Mean Squared Error (MSE), Mean Absolute Error (MAE), and R-squared to assess predictive performance \cite{azari2019bitcoin}.

These studies collectively highlight the growing importance of machine learning models in cryptocurrency price prediction. While traditional models like ARIMA offer simplicity for short-term forecasts, more advanced neural network-based models such as GRU and LSTM excel in handling long-term dependencies and highly volatile markets. These approaches provide valuable tools for investors navigating the complexities of cryptocurrency markets.

In addition, an research on Bitcoin mining hardware performance and lifespan highlighted practical considerations \cite{BrosMinerCrypto2024}. Application-Specific Integrated Circuits (ASIC) miners, which are specialized hardware designed exclusively for cryptocurrency mining, are known for their high efficiency and computational power. Reports by Cryptominerbros suggested that ASIC miners can operate reliably for 5--7 years and, with efficient cooling systems and maintenance, for over 10 years . Furthermore, the electricity transmission and distribution loss rate in South Korea remains around 3\%, as per KEPCO’s data \cite{Kim2022}. This research will build on such findings by directly linking surplus electricity to Bitcoin mining, providing a concrete analysis of its economic feasibility.

\subsection{Data Description}
This study utilizes various datasets related to surplus electricity power, Bitcoin mining, and price prediction. The key datasets include surplus electricity data from KEPCO, Bitcoin hash rate data, Bitcoin prices, and mining hardware capabilities, along with technical indicators. 

The Bitcoin hash rate represents the computational power required to generate coin blocks and indicates the mining difficulty. As the hash rate increases, the supply decreases, potentially leading to price increases \cite{kubal2022exploring}.

The surplus electricity power data from KEPCO spans monthly data from 2021 to 2023, which can be downloaded from the Public Data Portal \cite{KEPCO2024}. The data consists of regional information, including the total number of households equipped with solar power systems and the surplus electricity generated after net metering (measured in kWh). This study uses the surplus electricity data along with transmission loss rates to estimate the number of mining machines that can be operated and analyze their economic feasibility. Table~\ref{tab:surplus-power-stats} summarizes the surplus electricity data, including missing values, surplus power in kWh, and total households with solar power systems.

\begin{table}[ht]
\centering
\caption{Descriptive Statistics (Surplus Electricity Data for 2021-2023)}
\label{tab:surplus-power-stats}
\setlength{\tabcolsep}{5pt}
\begin{tabular}{l r r r r}
\toprule
\textbf{Variable} & \textbf{Mean} & \textbf{Max} & \textbf{Std. Dev.} \\
\midrule
Electricity Power (kWh) & 29,290 & 1,055,000 & 37,910 \\
Total Households   & 134.4 & 2,522 & 164.4 \\
\bottomrule
\end{tabular}
\end{table}

The "Electricity Power" variable represents the monthly surplus power in kilowatt-hours after net metering, with significant variability as indicated by the standard deviation. "Total Households" denotes the number of households equipped with solar power systems in each region, which also shows high variability across regions.

Bitcoin network data, including total network hash rate and daily average market prices, was sourced from public blockchain databases and cryptocurrency platforms (e.g., blockchain.com). The block reward (Bitcoin) you receive for mining once is also taken into account. Using these datasets, the daily Bitcoin mining capability and expected revenue for mining machines with specific hash rate are calculated. Table~\ref{tab:bitcoin-stats} presents descriptive statistics for Bitcoin-related data, including market price and hash rate.

\begin{table}[ht]
\centering
\caption{Descriptive Statistics (Bitcoin Data)}
\label{tab:bitcoin-stats}
\renewcommand{\arraystretch}{1.2}
\setlength{\tabcolsep}{5pt}
\begin{tabular}{l r r r r}
\toprule
\textbf{Variable} & \textbf{Mean} & \textbf{Max} & \textbf{Std. Dev.} \\ \midrule
Market Price (USD) & 16,940 & 67,560 & 16,310 \\
Hash Rate (TH/s) & 122,600,000 & 612,100,000 & 124,300,000 \\\bottomrule
\end{tabular}
\end{table}

"Market Price (USD)" reflects the daily average Bitcoin price in USD, showing high variability over the analysis period. “Hash Rate” indicates the computational power of the Bitcoin network, with significant fluctuations. For example, mining 1 BTC with a 500 TH/s machine like the Antminer S21+ would take over 7 years on average when mining solo, depending on network difficulty and total hash rate.

\section{Methodology}
We introduce the price prediction approaches considered in this study. Random Forest and LSTM are briefly introduced, and we refer to Breiman (2001) and Hochreiter (1997) for more detailed explanations \cite{breiman2001random}, \cite{Hochreiter1997}. In the following sections, we first discuss two machine learning methods, and then, describe the measures for model comparison.

\subsection{Random Forest Regressor}

Random Forest Regressor, proposed by Breiman (2001), is an ensemble learning method that combines multiple decision trees to produce robust regression estimates \cite{breiman2001random}. For a regression problem, let the training dataset be 
\begin{equation}
    \mathcal{D} = \{(x_i, y_i)\}_{i=1}^n,
\end{equation}
where $x_i \in \mathbb{R}^p$ is the $i$-th feature vector and $y_i \in \mathbb{R}$ is the corresponding target value. The Random Forest constructs an ensemble of $B$ decision tress, $\{T_b\}_{b=1}^B$, and the final prediction is obtained by averaging the outputs of all individual trees.

\subsubsection{Bootstrap Aggregating (Bagging)}
For each tree $T_b$ in the ensemble:
\begin{enumerate}
    \item A bootstrap sample $\mathcal{D}_b$ of size $n$ is drawn with replacement from $\mathcal{D}$. Due to sampling with replacement, each observation has a probability of approximately $P(\text{an observation is included in $\mathcal{D}_b$})=1-(1-\frac{1}{n})^n \approx 0.632.$
    \item The tree $T_b$ is then grown using the bootstrap sample $\mathcal{D}_b$.
\end{enumerate}

This procedure introduces diversity among the trees, which is essential for reducing the variance of the final model.

\subsubsection{Tree Construction}
A decision tree is structured as a series of connected points called \emph{nodes}. In this context, each node represents a decision point where the data is split into subsets based on a selected feature and threshold. Nodes that further split the data are called \emph{internal nodes}, while nodes that do not split any furthre are called \emph{leaf nodes}.
Within each decision tree $T_b$, the following steps are repeated recursively until a stopping criterion (e.g., minimum node size or maximum depth) is met:
\begin{enumerate}
    \item \textbf{Random Feature Selection:} At each internal node, rather than considering all $p$ features, a random subset of $m_{try}$ features is selected. A common choice for regression is:
    \begin{equation}
        m_{try} = \max(1, \lfloor p/3 \rfloor)
    \end{equation}
    This randomization helps to de-correlate the trees.
    \item \textbf{Optimal Split Determination:}  
For the selected subset of features at a given node, the algorithm searches for the optimal split. Consider a candidate split \(s\) that partitions the data at the node into two child nodes:
\[
L(s) = \{ i \mid x_i \text{ is assigned to the left child node under } s \},\]
\[R(s) = \{ i \mid x_i \text{ is assigned to the right child node under } s \}.\]
The optimal split \( s^* \) is chosen to minimize the sum of squared errors (SSE) in the child nodes:
\[
s^* = \argmin_{s} \left\{ \sum_{i \in L(s)} \left(y_i - \bar{y}_{L(s)}\right)^2 + \sum_{i \in R(s)} \left(y_i - \bar{y}_{R(s)}\right)^2 \right\},
\]
where
\[
\bar{y}_{L(s)} = \frac{1}{|L(s)|} \sum_{i \in L(s)} y_i, \quad \bar{y}_{R(s)} = \frac{1}{|R(s)|} \sum_{i \in R(s)} y_i.
\]
\end{enumerate}

Each tree is grown fully (or until a pre-specified stopping condition is met) without pruning. Although individual trees may overfit, averaging over many trees reduces the overall variance and improves generalization.

\subsubsection{Prediction}
Given a new observation $x \in R^p$, each tree $T_b$ provides a prediction $T_b(x)$. The final prediction of the Random Forest prediction is then the average of these predictions:

\begin{equation}
    \hat{f}_{RF}(x) = \frac{1}{B} \sum_{b=1}^B T_b(x).
\end{equation}
This averaging process leverages the "wisdom of the crowd" effect, resulting in a robust and stable prediction even if individual trees are noisy.

\subsection{Long Short-Term Memory}

Long Short-Term Memory is a specialized type of Recurrent neural network designed to capture long-term dependencies in sequential data. Traditional RNNs often struggle with vanishing or exploding gradients when processing long sequences \cite{Bengio1994}. LSTM overcomes these challenges by incorporating a sophisticated gating mechanism that regulates the flow of information, thereby enabling the network to maintain a stable gradient over extended time intervals.

Consider an input sequence \(\{x_t\}_{t=1}^{T}\) with \(x_t \in \mathbb{R}^{d}\) representing the input at time \(t\). The LSTM cell maintains two key state vectors at each time step \(t\):
\begin{itemize}
    \item the \textbf{hidden state} \(h_t \in \mathbb{R}^{h}\), which serves as the output of the cell, and
    \item the \textbf{cell state} \(C_t \in \mathbb{R}^{h}\), which functions as the internal memory of the cell.
\end{itemize}

At each time step, the LSTM cell performs the following computations:

\subsubsection{Forget Gate}  
The forget gate determines which information from the previous cell state \(C_{t-1}\) should be discarded. It is defined as:
\[
f_t = \sigma\big(W_f x_t + U_f h_{t-1} + b_f\big),
\]
where:
\begin{itemize}
    \item \(W_f \in \mathbb{R}^{h \times d}\) and \(U_f \in \mathbb{R}^{h \times h}\) are weight matrices,
    \item \(b_f \in \mathbb{R}^{h}\) is the bias vector, and
    \item \(\sigma(\cdot)\) denotes the sigmoid activation function, outputting values between 0 and 1.
\end{itemize}

\subsubsection{Input Gate and Candidate Memory}
The input gate controls the incorporation of new information into the cell state. It comprises two parts:
\[
i_t = \sigma\big(W_i x_t + U_i h_{t-1} + b_i\big),
\]
\[
\tilde{C}_t = \tanh\big(W_C x_t + U_C h_{t-1} + b_C\big),
\]
where:
\begin{itemize}
    \item \(W_i,\, W_C \in \mathbb{R}^{h \times d}\) and \(U_i,\, U_C \in \mathbb{R}^{h \times h}\) are weight matrices,
    \item \(b_i,\, b_C \in \mathbb{R}^{h}\) are bias vectors, and
    \item \(\tanh(\cdot)\) is the hyperbolic tangent function, scaling the candidate values to lie between \(-1\) and \(1\).
\end{itemize}

\subsubsection{Cell State Update}  
The new cell state \(C_t\) is computed by combining the previous cell state, modulated by the forget gate, with the candidate memory scaled by the input gate:
\[
C_t = f_t \odot C_{t-1} + i_t \odot \tilde{C}_t,
\]
where \(\odot\) denotes element-wise multiplication.

\subsubsection{Output Gate and Hidden State}  
Finally, the output gate determines the hidden state \(h_t\), which is used both for output and for the next time step's computations:
\[
o_t = \sigma\big(W_o x_t + U_o h_{t-1} + b_o\big),
\]
\[
h_t = o_t \odot \tanh(C_t),
\]
with:
\begin{itemize}
    \item \(W_o \in \mathbb{R}^{h \times d}\), \(U_o \in \mathbb{R}^{h \times h}\), and \(b_o \in \mathbb{R}^{h}\) being the corresponding weight matrices and bias vector.
\end{itemize}

\begin{figure}[ht]
\centering
\includegraphics[width=0.7\columnwidth]{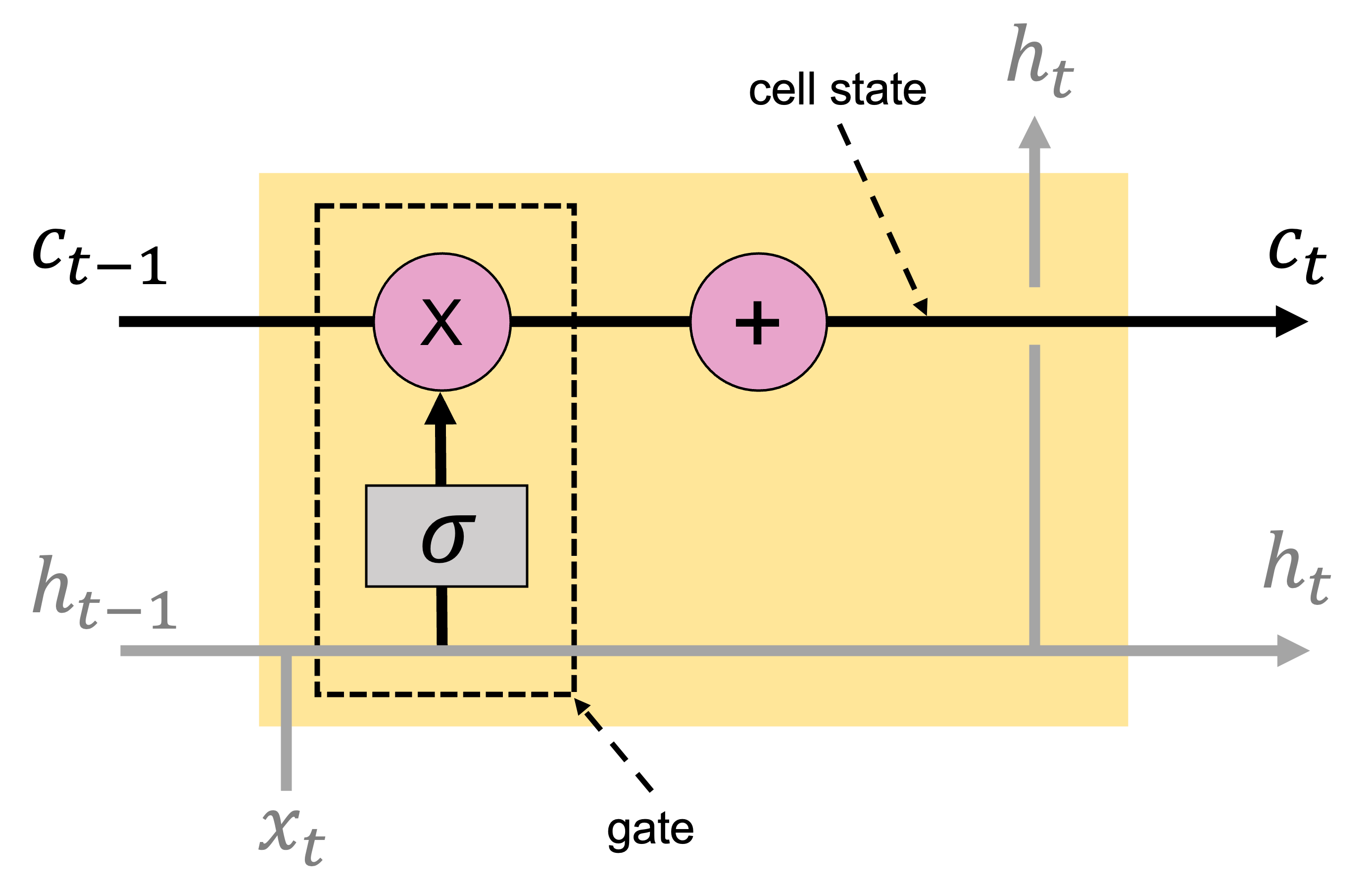}
\caption{Cell State in an LSTM}
\label{fig:lstm-cell-state}
\end{figure}

Figure~\ref{fig:lstm-cell-state} illustrates the internal structure of an LSTM cell, highlighting the cell state (depicted as a horizontal conveyor belt) and the gate that modulate the flow of information.

The entire LSTM network is trained using Backpropagation through time, adjusting the parameters
\[
\{W_f,\, U_f,\, b_f,\, W_i,\, U_i,\, b_i,\, W_C,\, U_C,\, b_C,\, W_o,\, U_o,\, b_o\}
\]
to minimize a predefined loss function.

By selectively retaining or discarding information at each time step, LSTM networks are particularly effective for modeling time-series data and other sequential tasks that require learning long-term dependencies.

\subsection{Model Comparison}
Performance is evaluated using metrics like Mean Squared Error and Mean Absolute Error.

\subsubsection{Mean Squared Error}
The Mean Squared Error measures the average squared difference between the predicted values and the actual values. It is mathematically defined as:
\[
\text{MSE} = \frac{1}{N} \sum_{i=1}^{N} (y_i - \hat{y}_i)^2
\]
where \( y_i \) represents the actual value, \( \hat{y}_i \) represents the predicted value, and \( N \) is the number of data points. Since MSE squares the errors, it gives more weight to larger errors, making it sensitive to outliers.

\subsubsection{Mean Absolute Error}
The Mean Absolute Error computes the average absolute difference between the predicted values and the actual values. It is mathematically defined as:
\[
\text{MAE} = \frac{1}{N} \sum_{i=1}^{N} |y_i - \hat{y}_i|
\]
Unlike MSE, MAE uses absolute differences, treating all errors equally and being less sensitive to outliers compared to MSE.

\section{Data Analysis}
Before proceeding with the analysis, several key assumptions are made:
\begin{enumerate}
    \item The location of Bitcoin miners is not considered.
    \item The availability of surplus power is assumed to be 24 hours per day.
    \item A block is mined every 10 minutes \cite{traders2025}.
\end{enumerate}

\subsection{Bitcoin Price}
\begin{figure}[ht]
\centering
\includegraphics[width=\columnwidth]{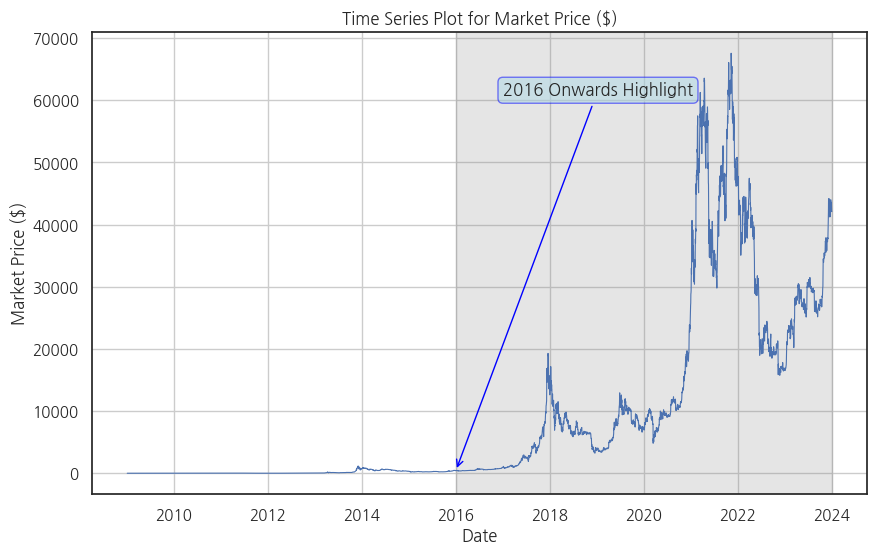}
\caption{Bitcoin Price Data Over the Entire Period}
\label{fig:bitcoin-price-all}
\end{figure}

Figure~\ref{fig:bitcoin-price-all} illustrates Bitcoin price data over the entire observed period. As seen in the figure, price fluctuations prior to 2016 are negligible. Therefore, this analysis focuses on data from 1, January, 2016 to 23, September, 2023 for Bitcoin price prediction, which is indicated by the gray background in the figure.

\subsection{Surplus Electricity Power}
\begin{figure}[ht]
    \centering
    \includegraphics[width=\linewidth]{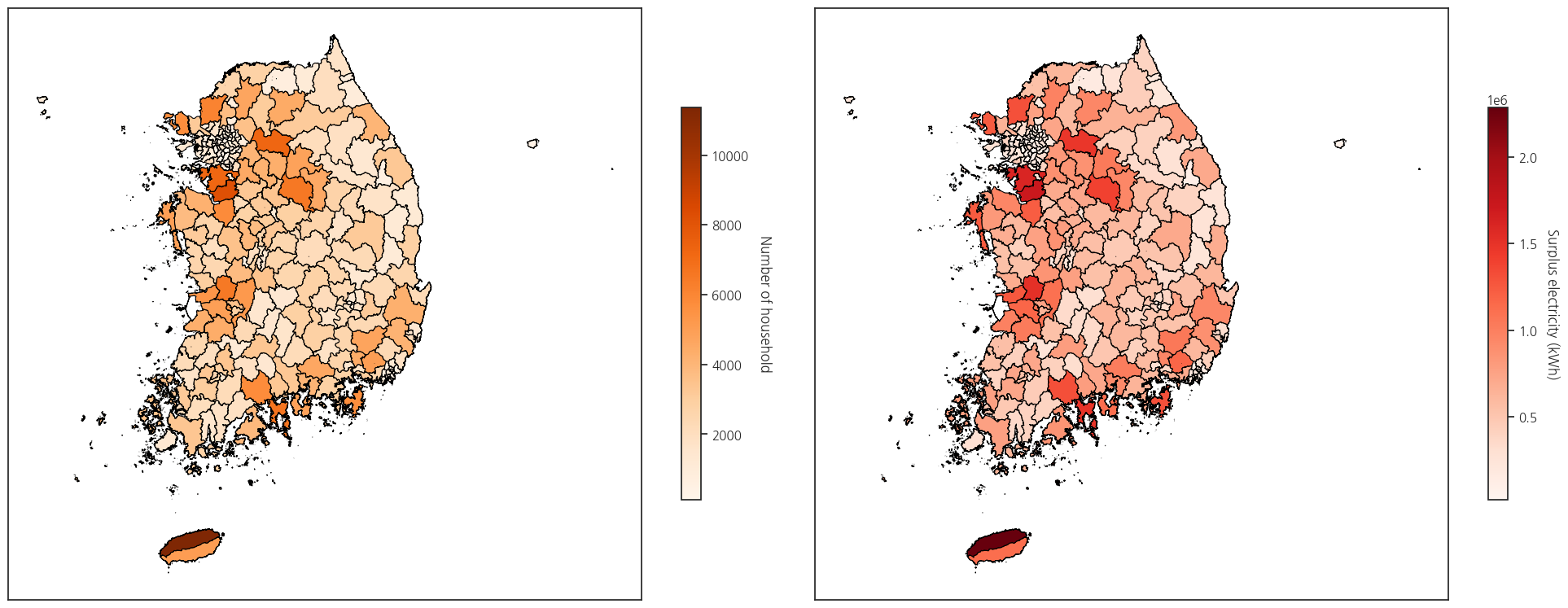}
    \caption{Average number of households and average surplus electricity (kWh) by region.}
    \label{fig:avg_surplus_power}
\end{figure}

Figure~\ref{fig:avg_surplus_power} presents a graph showing the average number of households applying for solar power net metering and the average surplus electricity power (kWh) over a period of three years. Darker colors indicate higher values.

High levels of non-metered power generation were observed in regions near the southern coastal area, Jeju Island, and the western coastline.

\begin{figure}[ht]
    \centering
    \includegraphics[width=\linewidth]{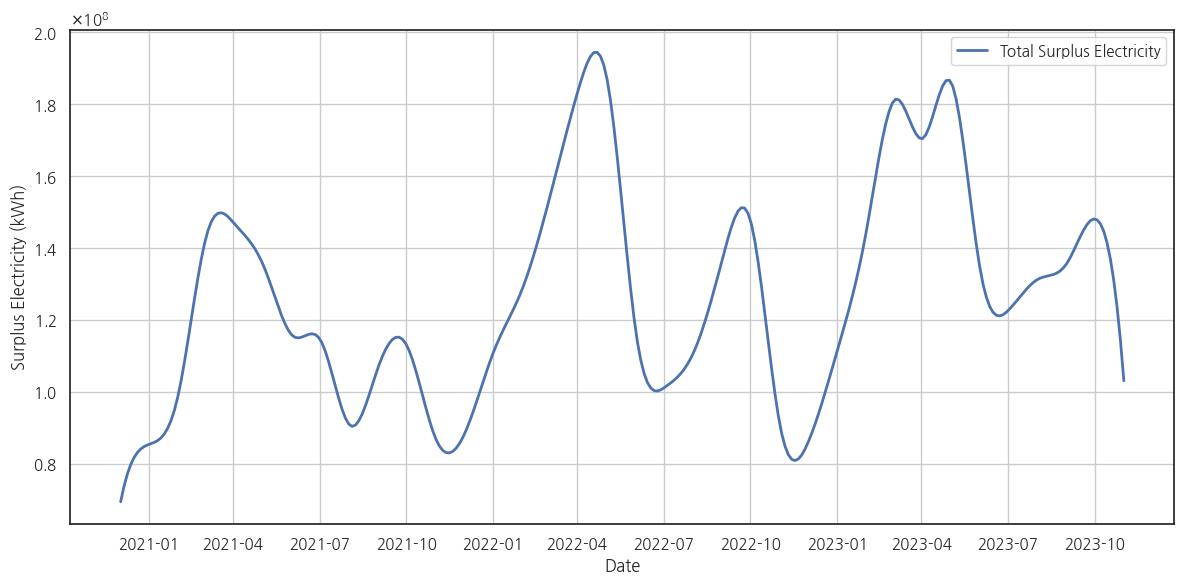}
    \caption{Monthly total surplus power over three years}
    \label{fig:monthly_surplus_power}
\end{figure}

 Figure~\ref{fig:monthly_surplus_power} shows the total excess power generation aggregated over three years, displayed on a monthly basis. It can be observed that the highest excess power generation values occur around May of each year. This indicates a clear seasonal effect on surplus power.

\subsection{Development of Independent Variables}
In Pabuccu et al. (2023), the authors selected independent variables for modeling, as shown in Table~\ref{tab:independent_variables}, to identify short-term Bitcoin price trend \cite{pabucccu2023forecasting}. Short-term moving average indicators, such as moving average (MA) and weighted moving average (WMA), place greater weight on the most recent observations to capture speculative volatility, while Momentum and stochastic indicators (K\% and D\%) evaluate the strength and sustainability of the current trend. The Relative Strength Index (RSI) identifies overbought and oversold levels to assess market conditions, aiming to improve model performance. Similarly, in this analysis, independent variables for Bitcoin price prediction were developed, as summarized in Table~\ref{tab:independent_variables}.

\begin{table}[ht]
\centering
\caption{Development of Independent Variables}
\label{tab:independent_variables}
\renewcommand{\arraystretch}{1.2}
\begin{tabular}{>{\centering\arraybackslash} p{2cm} m{6cm}}
\toprule
\multicolumn{1}{c}{\textbf{Variable Name}} & \multicolumn{1}{c}{\textbf{Definition}} \\ \midrule
SMA\_14 &The average price over a 14-day period.\\
\midrule
WMA\_14 & Weighted moving average price over 14 days, assigning higher weights to more recent prices. \\
\midrule
Momentum & 
\( P_\text{today} - P_{N\text{ days ago}} \); \( P_i\) is the price on day i and \(N = 1\) in this analysis. Represents the speed and direction of price change. Positive values indicate upward momentum, negative values indicate downward momentum, and it is used to assess the sustainability of price trends. \\
\midrule
K\% & 
\( \frac{P_\text{today} - L_{14}}{H_{14} - L_{14}} \times 100 \), where \(H_{14}\) is the highest price over the past 14 days and \(L_{14}\) is the lowest price over the past 14 days. \\
\midrule
D\% &3-day simple moving average of K\%.\\
\midrule
RSI & 
\( 100 - \frac{100}{1 + \text{RS}} \), where \(\text{RS} = \frac{\text{Average Gain over N days}}{\text{Average Loss over N days}}\). Measures the relative strength of price gains and losses, ranging between 0 and 100. RSI values above 70 indicate overbought conditions, while values below 30 indicate oversold conditions. \\
\bottomrule
\end{tabular}
\end{table}

\section{Empirical Analysis}
\subsection{Bitcoin Price Prediction}
\subsubsection{Model Development}
Following prior research, two predictive models, Random Forest and LSTM, were employed for Bitcoin price prediction. The dataset was divided into a training set (2016/01/16–2022/12/31) and a test set (2023/01/01–2023/12/31). 

For the LSTM model, training was conducted with a 16GB RAM M2 chip, using 20 epochs, which can be changed in future studies with extended time allocated for model tuning.

\subsubsection{Modeling Results}
The prediction results of the two models are shown in Table~\ref{tab:prediction_results}. The Random Forest model achieved an \( R^2 \) of 0.91, higher than the LSTM model's 0.85. For MSE and MAE, LSTM had a slightly lower value, the LSTM model showed a marginally better performance.

\begin{table}[ht]
\centering
\renewcommand{\arraystretch}{1.2}
\caption{Bitcoin Price Prediction Results}
\label{tab:prediction_results}
\begin{tabular}{c ccc}
\toprule
& \multicolumn{3}{c}{Results} \\
\cmidrule(lr){2-4}
Model & MAE & MSE & \textbf{\( R^2 \)} \\
\midrule
\textbf{Random Forest} & 1,329 & 2,909,857 & 0.91 \\
\textbf{LSTM}          & 1,736 & 4,643,802 & 0.85 \\ 
\bottomrule
\end{tabular}
\end{table}

\begin{figure}[ht]
    \centering
    \includegraphics[width=\linewidth]{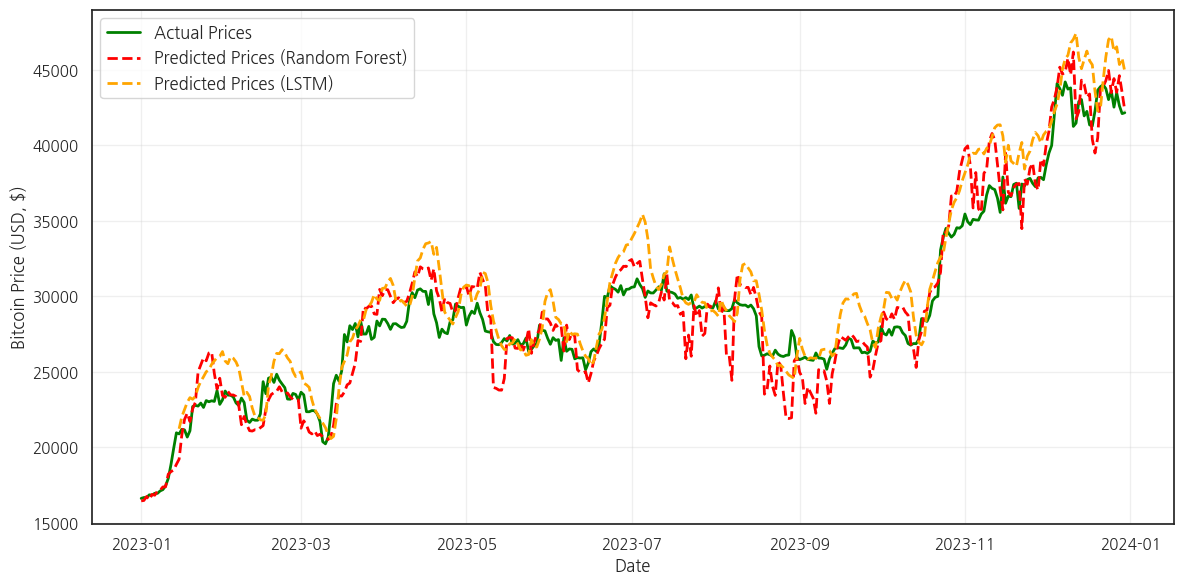}
    \caption{Comparison the predicted Bitcoin prices with actual prices using the Random Forest and the LSTM}
    \label{fig:rf_vs_lstm_vs_actual}
\end{figure}

Figure~\ref{fig:rf_vs_lstm_vs_actual} compares the predicted Bitcoin prices with actual prices (green solid line) using the Random Forest model (red dashed line) and the LSTM model (yellow dashed line).

\noindent
As observed in the graph, both models closely track the actual prices, except for instances of sharp price drops. The LSTM model utilized in this analysis took a sliding window approach, where the previous 14 days of data are used as inputs to predict the target variable. Consequently, the initial 14 days are absent in the visualization of predicted values, as they represent the input-only phase of the model. This is a natural limitation of models that rely on past sequences to generate forecasts.

\subsection{Estimated Number of Mined Bitcoins}
\subsubsection{Mining Equipment}
For the analysis, the most advanced Bitcoin mining equipment available as of November 2024 was assumed.

\begin{figure}[ht]
    \centering
    \includegraphics[width=4cm]{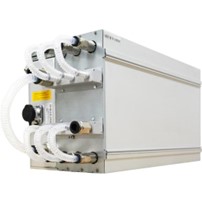}
    \caption{Bitcoin Miner S21 XP Hyd. \cite{BITMAIN2024}}
    \label{fig:miner_photo}
\end{figure}

Figure~\ref{fig:miner_photo} shows an actual image of the Bitcoin Miner S21 XP Hyd. from BITMAIN \cite{BITMAIN2024}. This miner is based on the SHA-256 algorithm and features water cooling technology, offering both high performance and energy efficiency. It achieves a maximum hash rate of 473 TH/s (the number of calculations the equipment can perform per second), a power consumption of 5,676 W, and a power efficiency of 12 J/TH, representing the highest performance currently available.

\subsubsection{Number of Available Miners per Month}
\begin{figure}[ht]
    \centering
    \vspace{-1em}
    \includegraphics[width=\linewidth]{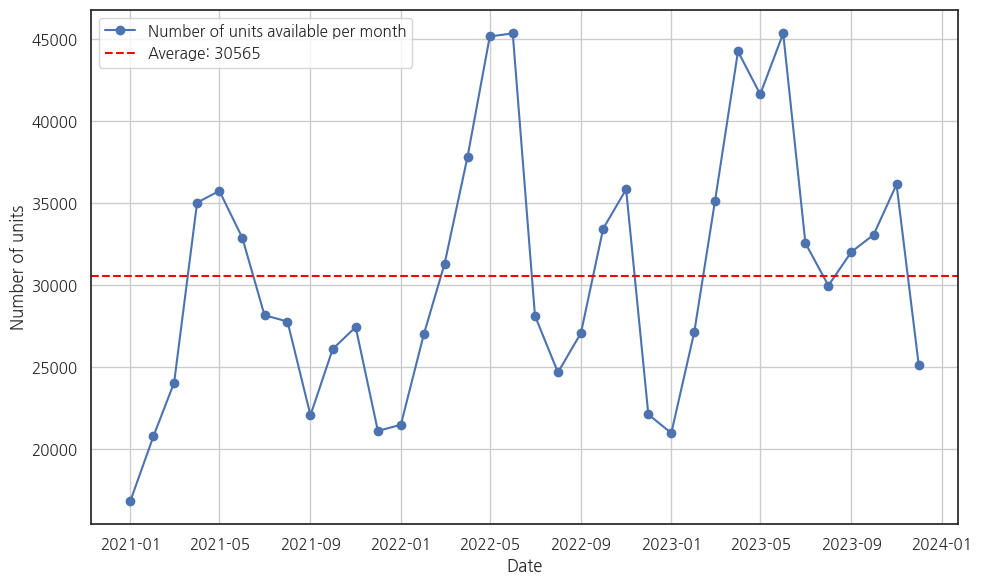}
    \vspace{-2em}
    \caption{Number of units available per month}
    \vspace{-1.5em}
    \label{fig:monthly_miners}
\end{figure}

Assuming the excess electricity is utilized to operate the mining equipment, the number of miners that can operate on a monthly basis was calculated. In this study, the power transmission loss rate was assumed to be 3.59\%, based on the 10-year average (2012–2021) reported by KEPCO \cite{Kim2022}. After accounting for this loss, the number of operational miners per month was calculated, as shown in Figure~\ref{fig:monthly_miners}. On average, 30,565 miners can operate monthly over three years, with a maximum of 45,439 miners.

\subsubsection{Expected Number of Bitcoins Mined}
As shown in Table~\ref{tab:block_reward}, Bitcoin undergoes a halving event approximately every four years, reducing the block reward (the number of Bitcoins generated per mined block) by half. During the analysis period (2021–2023), the block reward remained at 6.25 BTC (valid from May 12, 2020, to April 20, 2024) since there were no halving events within this period.

\begin{table}[ht]
\renewcommand{\arraystretch}{1.2}
\centering
\caption{Bitcoin Halving Events and Block Rewards}
\label{tab:block_reward}
\begin{tabular}{r r}
\toprule
\multicolumn{1}{c}{\textbf{Date}} & \multicolumn{1}{c}{\textbf{Block Reward (BTC)}} \\ \midrule
November 28, 2012     & 25 BTC \\ \midrule
July 10, 2016         & 12.5 BTC \\ \midrule
May 12, 2020          & 6.25 BTC \\ \midrule
April 20, 2024        & 3.125 BTC \\ \bottomrule
\end{tabular}
\end{table}

Next, two simulation scenarios were considered for the operational number of miners. As shown in Table~\ref{tab:simulation_settings}, the first simulation used the monthly number of available miners as-is, with a maximum of 45,439 miners. While this allows for maximum utilization of excess electricity, it also results in idle miners during months with lower excess electricity.

\begin{table}[ht]
\renewcommand{\arraystretch}{1.2}
\centering
\caption{Simulation Settings for Operational Miners}
\label{tab:simulation_settings}
\begin{tabular}{>{\centering\arraybackslash} p{1.6cm} m{6.4cm}}
\toprule
\multicolumn{1}{c}{\textbf{Simulation}} & \multicolumn{1}{c}{\textbf{Details}} \\ \midrule
Simulation 1        & Based on the monthly number of available miners. Maximum: 45,439 miners. \\ \midrule
Simulation 2        & Based on the average operational miners over three years (30,565 miners). This reduces idle equipment and initial purchase costs. \\ \bottomrule
\end{tabular}
\end{table}

\begin{figure}[ht]
    \centering
    \includegraphics[width=\linewidth]{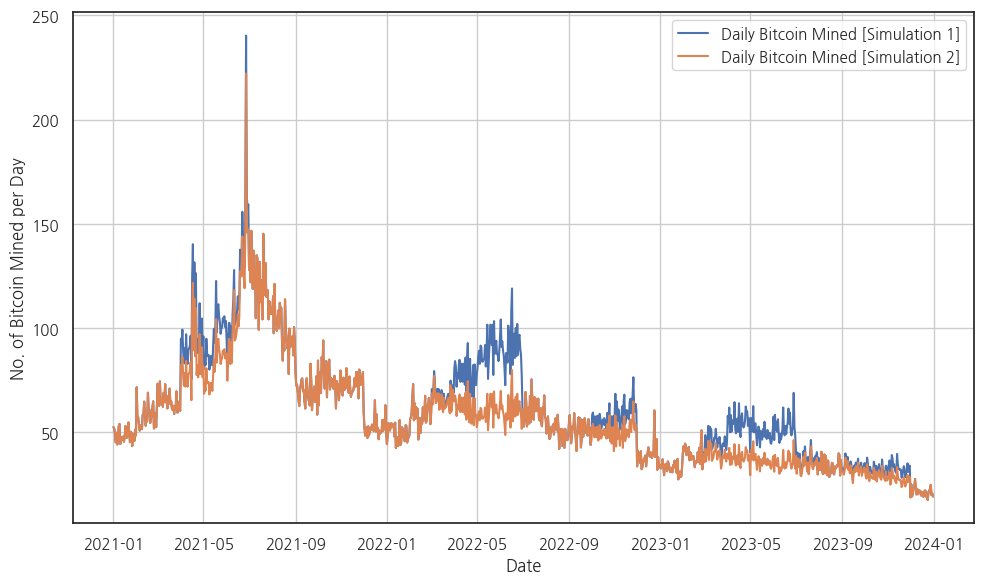}
    \caption{Expected Daily Number of Bitcoins Mined (By Simulation)}
    \label{fig:daily_bitcoin_mined}
\end{figure}

In Simulation 2, the number of miners was fixed at 30,565, the three-year average, to minimize idle equipment and reduce initial costs. Monthly data were converted to daily averages for analysis. Figure~\ref{fig:daily_bitcoin_mined} shows the expected daily number of Bitcoins mined for each simulation. 

As expected, higher expected numbers of mined Bitcoins are observed in May 2022 and May 2023 for Simulation 1 compared to Simulation 2.

\subsection{Cost Calculation}
Table~\ref{tab:cost_items} lists the items considered in cost calculation and items excluded from the analysis. The largest cost is the mining equipment, with each Bitcoin Miner S21 XP Hyd. priced at \$10,165 per unit. For conservative calculations, a lifespan of 7.5 years was assumed, with a straight-line monthly depreciation applied \cite{BrosMinerCrypto2024}. 

The second item, transmission and distribution loss, was already considered when estimating the number of operational miners.

\begin{table}[ht]
\renewcommand{\arraystretch}{1.2}
\centering
\caption{Cost Considerations and Excluded Items}
\label{tab:cost_items}
\begin{tabular}{>{\centering\arraybackslash} m{1.5cm} m{6.5cm}}
\toprule
\multicolumn{1}{c}{\textbf{Category}} & \multicolumn{1}{c}{\textbf{Details}}\\ \midrule
Mining Equipment Cost & 
Bitcoin Miner S21 XP Hyd. (BITMAIN)
\begin{itemize}
    \item \$10,165 per unit
    \item 7.5 years (90 months) for lifespan
    \item Assume monthly straight-line method for depreciation
\end{itemize} \\ \midrule
Transmission Loss & 
Transmission loss rate
\begin{itemize}
    \item 3.59\%
    \item Based on a 10-year average reported by Korea Electric Power Corporation
\end{itemize} \\ \midrule
Excluded Items &
Networking Equipment
\begin{itemize}
    \item High-speed routers, switches, cables
\end{itemize}
Replacement Costs
\begin{itemize}
    \item Fans, power supplies, chips, and repair costs
\end{itemize}
Labor Costs
\begin{itemize}
    \item IT engineers and cooling system managers
\end{itemize}
Profit Share or Surplus Electricity Providers, Cooling System, etc.
\\ \bottomrule
\end{tabular}
\end{table}

The profit share for surplus electricity providers, while an important consideration for future studies, is not defined in this analysis.

\subsection{Profit Calculation by Simulation}
Using the expected daily mining volume (Section~4.2) and cost calculation (Section~4.3), the daily profit for each simulation was calculated. Initially, profits were computed without deducting the cost of mining equipment. Our approach focuses on the miner's proportional contribution to the total network hash rate and the corresponding predicted market price.

The daily mining revenue is calculated by multiplying the predicted Bitcoin price with the estimated number of Bitcoins mined per day. Specifically, the number of Bitcoins mined per day is given by the product of:
\begin{enumerate}
    \item the fixed block reward (6.25 BTC, as detailed in Section~4.2.3),
    \item the proportion of the total network hash rate attributed to the simulated mining operation, and
    \item the average number of blocks mined per day (144, given that a block is mined every 10 minutes as stated in Assumption 3 in Section~3).
\end{enumerate}

This calculation can be expressed as:

\begin{equation}
\begin{aligned}
& \text{Daily Mining Revenue} \\
& = \text{Predicted Price} \times \text{BTC/day} \nonumber\\
& = \text{Predicted Price} \times \text{Block Reward} \times \frac{H}{H_{net}} \times 144
\end{aligned}
\end{equation}

where:
\begin{itemize}
    \item \(\text{Predicted Price}\) is the market price of Bitcoin predicted by the simulation (which may vary between simulation runs),
    \item \(\text{Block Reward}\) is the fixed reward of 6.25 BTC per block,
    \item \(\frac{H}{H_{\text{net}}}\) represents the fraction of the network's total hash rate employed by the simulated miner, \( H \) is the total hash rate simulated by the user (sum of miner capacities), and \( H_{\text{net}} \) is the total Bitcoin network hash rate.
    \item \(144\) is the number of blocks mined per day, based on an average block time of 10 minutes.
\end{itemize}

This formulation directly ties the simulated miner's computational power share to its expected daily revenue. Our approach is solely designed to compare profitability under different simulation scenarios. The focus is on assessing how variations in the predicted Bitcoin price and the hash rate affect the revenue.

\begin{figure}[ht]
    \centering
    \includegraphics[width=\linewidth]{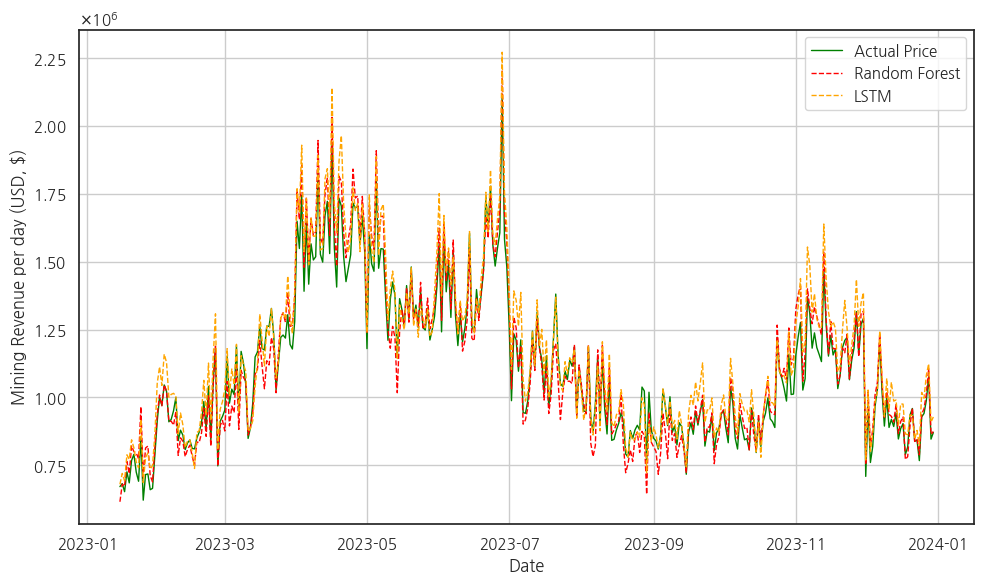}
    \caption{Daily Mining Profit (Before Deducting Costs) - Simulation 1}
    \label{fig:daily_profit_sim1}
\end{figure}

\begin{figure}[ht]
    \centering
    \includegraphics[width=\linewidth]{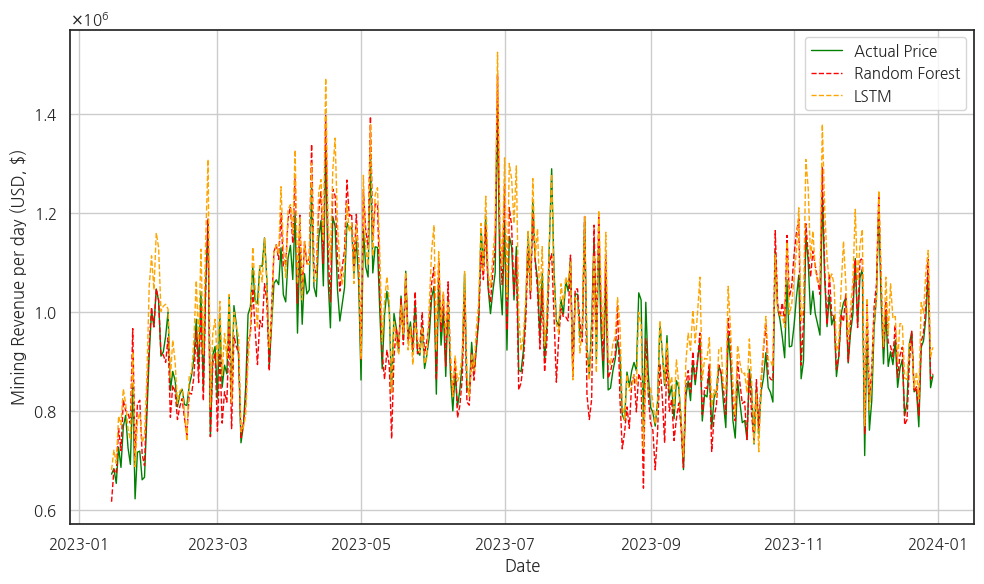}
    \caption{Daily Mining Profit (Before Deducting Costs) - Simulation 2}
    \label{fig:daily_profit_sim2}
\end{figure}

Figures~\ref{fig:daily_profit_sim1} and~\ref{fig:daily_profit_sim2} show the daily mining profit for Simulation 1 and Simulation 2, respectively. The graphs demonstrate that the actual values and model predictions are closely aligned, indicating reliable forecasting.

\begin{figure}[ht]
    \centering
    \includegraphics[width=\linewidth]{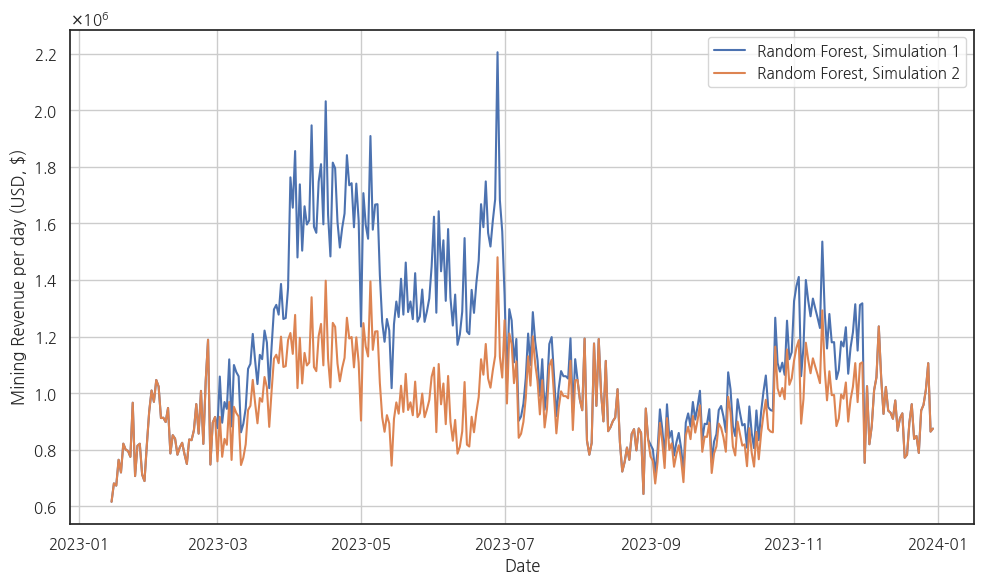}
    \caption{Daily Mining Profit (Before Deducting Costs) - Random Forest}
    \label{fig:daily_profit_rf}
\end{figure}

To analyze the situation for each simulation using a single model, the Random Forest model was visualized in Figure~\ref{fig:daily_profit_rf}. As expected, Simulation 1, which operates at maximum capacity, showed significantly higher profits from April to July compared to Simulation 2.

\subsection{Final Profit Comparison by Model and Simulation}

The analysis includes four cases based on the models and simulations used. Table~\ref{tab:final_profit_comparison} presents the final profit comparison including revenue and cost calculations over 12 months of operation (January to December 2023) for each case. In the table, ‘Sim’ in the Case column is short for 'Simulation', and M denotes millions of USD.

\begin{table}[ht]
\centering
\caption{Final Profit Comparison by Model and Simulation}
\label{tab:final_profit_comparison}
\begin{tabular}{r l r r}
\toprule
\multicolumn{1}{c}{\textbf{Case}}
    & \multicolumn{1}{c}{\textbf{Revenue (M)}} 
    & \multicolumn{1}{c}{\textbf{Cost (M)}} 
    & \multicolumn{1}{c}{\textbf{Profit (M)}}\\ \midrule
Actual Price, Sim 1
    & \$390
    & \$62
    & \$328 \\ \midrule
Actual Price, Sim 2 
    & \$333
    & \$41
    & \$292 \\ \midrule
Random Forest, Sim 1 
    & \$393 (+0.8\%)
    & \$62
    & \$331\\ \midrule
Random Forest, Sim 2 
    & \$335 (+0.6\%)
    & \$41
    & \$294 \\ \midrule
LSTM, Sim 1 
    & \$409 (+4.9\%)
    & \$62
    & \$348 \\ \midrule
LSTM, Sim 2 
    & \$349 (+4.8\%)
    & \$41
    & \$308 \\ \bottomrule
\end{tabular}
\end{table}

In the table, the percentages in parentheses next to each model's revenue indicate the relative difference from the actual price scenario. Notably, the case with the lowest profit occurs when the Bitcoin price is predicted using the Random Forest model under Simulation 2 (which assumes 30,565 miners), resulting in a profit of approximately 294 million USD over 12 months—equivalent to over 24.5 million USD in revenue per month.

The cost calculation in Table~\ref{tab:final_profit_comparison} is based on the number of miners, the unit price of each Bitcoin miner, the operating period (12 months), and the assumed lifetime of a miner (7.5 years, equivalent to 90 months). Specifically, the cost for each simulation is computed as follows:
\[
\text{Cost} = \frac{\text{Number of Miners} \times \text{Miner Price} \times 12\ (\text{months})}{90\ (\text{months})}.
\]

For Simulation 1, using 40,439 miners at a unit price of \$10,165, the cost is:
\[
\frac{40439 \times 10165 \times 12}{90} \approx 62\text{M USD},
\]
and for Simulation 2, using 30,565 miners, the cost is:
\[
\frac{30565 \times 10165 \times 12}{90} \approx 41\text{M USD}.
\]

\section{Conclusion}
\begin{figure}[ht]
  \centering
  \begin{tikzpicture}[
    node distance=1cm,
    every node/.style={font=\small, align=center},
    startstop/.style={rectangle, rounded corners, draw, fill=red!20, minimum width=4cm, minimum height=0.8cm},
    process/.style={rectangle, draw, fill=blue!20, minimum width=4cm, minimum height=0.8cm},
    arrow/.style={-{Stealth}, thick}
  ]
    % Nodes
    \node (input) [startstop] {Load Data:\\ Monthly Surplus Energy (2021--2023)\\ Daily Bitcoin Price \& Hash Rate (2021--2023)};
    \node (pre)   [process, below=of input]  {Data Cleaning \& Alignment};
    \node (pred)  [process, below=of pre]    {Price Prediction:\\ Actual (Baseline)\\ LSTM Model\\ Random Forest Regressor};
    \node (param) [process, below=of pred]   {Simulation Parameters:\\ Max Units: 45.439 Antminers\\ Avg Units: 30.565 Antminers};
    \node (cost)  [process, below=of param]  {Deduct Necessary Expenses:\\ Power Loss Rate, Miner Purchase Cost};
    \node (out)   [startstop, below=of cost]   {Compute Net Revenue:\\ For Each Model \& Simulation};

    % Arrows
    \draw [arrow] (input) -- (pre);
    \draw [arrow] (pre) -- (pred);
    \draw [arrow] (pred) -- (param);
    \draw [arrow] (param) -- (cost);
    \draw [arrow] (cost) -- (out);
    
  \end{tikzpicture}
  \caption{Analysis workflow}
  \label{fig:analysis_flow_vertical_expenses}
\end{figure}
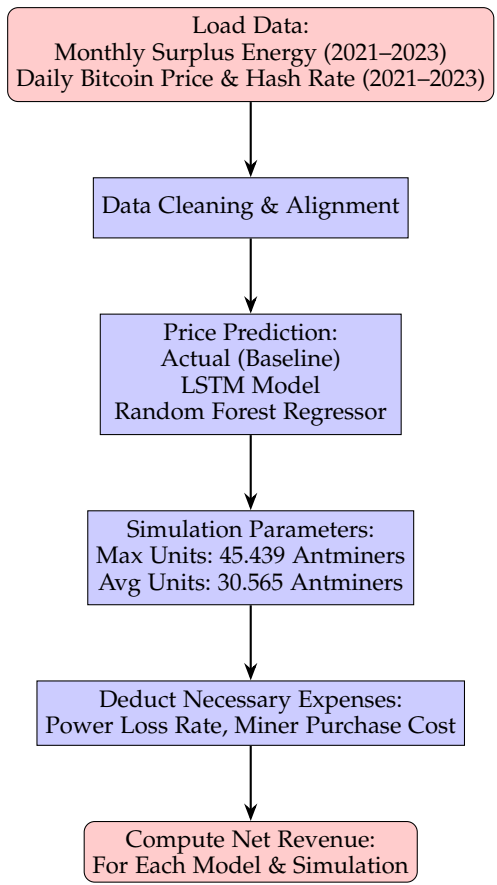

This study analyzed the feasibility of utilizing surplus electricity generated from solar power for Bitcoin mining as a way to resolve the debt problem of KEPCO and improve the efficiency of surplus electricity utilization. Based on the performance of the latest mining equipment, the number of Bitcoins that could be mined was calculated and the profitability was evaluated considering the hash rate of the Bitcoin network. Furthermore, Random Forest and LSTM models were utilized to predict Bitcoin price volatility, and their prediction accuracy was compared. Random Forest model showed relatively higher performance in terms of MSE and MAE. The simulation results, which indicated profit differences across various scenarios, demonstrate that even under conservative estimates, the additional revenue generated from Bitcoin mining can be significant. Figure ~\ref{fig:analysis_flow_vertical_expenses} is a simplified schematic representation of the process of this empirical study.

The profitability analysis shows that utilizing surplus electricity for Bitcoin mining can minimize power losses, mitigate KEPCO's unsettled payment issues, and generate economic profit. This represents the approach to converting unused electricity resources into additional income, which offers significant implications for enhancing the efficient use of energy resources at a national level.

Bitcoin is increasingly recognized as a leading digital asset in the global financial system, and its value is reflected in its price. Recent movements by major nations, including the United States, to adopt Bitcoin as a national asset or explore its utilization are noteworthy. This suggests that Bitcoin could serve as more than just an investment vehicle, potentially playing the role of a safe haven asset, and its role in the digital economy is expected to expand further.

However, in order for the proposed approach to be realized, legal and institutional barriers must first be addressed. Currently, KEPCO is only allowed to engage in electricity sales under the KEPCO Act (Article 13), making direct entry into new businesses such as Bitcoin mining impossible \cite{KEPCOAct2022}. Therefore, legislative amendments and policy support must precede the use of surplus electricity for mining. This could pave the way for innovating the existing structure of the power industry and converting unused electricity resources into new economic value.

This study is significant as it is the first empirical analysis in South Korea to evaluate the economic feasibility of linking KEPCO's surplus electricity with Bitcoin mining. It highlights the potential to achieve efficient use of unused electricity, alleviate KEPCO's debt, and position Bitcoin mining as a new revenue-generating tool for the national economy.

Although this study empirically analyzed the economic feasibility of using surplus solar electricity for Bitcoin mining, several limitations and future research directions remain. Identifying optimal regions with abundant surplus electricity and minimal power losses is essential, which requires a comprehensive evaluation of factors such as proximity to substations, stability of power supply, and impacts on local economies to select the best locations for mining facilities. Additionally, technological improvements should be explored to efficiently integrate solar power generation and Bitcoin mining systems, with a focus on coupling with energy storage systems and optimizing cooling systems to maximize mining equipment's energy efficiency. Since the Bitcoin hash rate is the most critical factor influencing mining profitability, future research should emphasize forecasting hash rate trends to refine profitability predictions. Furthermore, extending the methodology presented in this study to other renewable energy sources, such as wind and hydroelectric power, represents an important avenue to diversify the use of energy.

\ifCLASSOPTIONcaptionsoff
  \newpage
\fi

\bibliographystyle{IEEEtran}
\bibliography{ref}

\end{document}